\documentclass[prb]{revtex4}
\begin{document}

\title
{Order parameter phase locking as a cause of a zero bias peak in
the differential tunneling conductance of bilayers with
electron-hole pairing}
\author{A.\,I.\,Bezuglyj}
\email{bezugly@ic.kharkov.ua} \affiliation{NSC Kharkov Institute
of Physics and Technology, 61108, Kharkov, Ukraine}
\author{S.\,I.\,Shevchenko}
\email{shevchenko@ilt.kharkov.ua}
\affiliation{%
B.\,I.\, Verkin Institute for Low Temperature Physics and
Engineering National Academy of Sciences of Ukraine,  Kharkov
61103, Ukraine}

\begin{abstract}
In $n-p$ bilayer systems an exotic phase-coherent state emerges
due to Coulomb pairing of $n$-layer electrons with $p$-layer
holes. Unlike Josephson junctions, the order parameter phase may
be locked by matrix elements of interlayer tunneling in $n-p$
bilayers. Here we show how the phase locking phenomenon specifies
the response of the electron-hole condensate to interlayer
voltages. In the absence of an applied magnetic field, the phase
is steady-state (locked) at low interlayer voltages, $V<V_c$,
however the phase increases monotonically with time (is unlocked)
at $V>V_c$. The change in the system dynamics at $V=V_c$ gives
rise to a peak in the differential tunneling conductance. The peak
width $V_c$ is proportional to the absolute value of the tunneling
matrix element $|T_{12}|$, but its height does not depend on
$|T_{12}|$; thus the peak is sharp for small $|T_{12}|$. A
sufficiently strong in-plane magnetic field reduces considerably
the peak height. The present results are in qualitative agreement
with the zero bias peak behavior that has recently been observed
in bilayer quantum Hall ferromagnets with spontaneous interlayer
phase coherence.
\end{abstract}


\maketitle

 The idea that in bilayer $n-p$ structures consisting of an electron-
conductivity layer  ($n$-layer) and a hole-conductivity layer
($p$-layer) the Coulomb attraction of electrons and holes may lead
a formation of  electron-hole pairs with spatially separated
components was put forward rather long ago [1, 2]. As a
result of Bose-Einstein condensation of these pairs, there arises
a peculiar superfluid (phase-coherent) state, in which a
dissipationless motion of pairs gives rise to equal-in-magnitude
and oppositely directed supercurrents in $n$- and $p$-layers. At
present, two variants of the systems have been realized
experimentally, where an excitonic condensate with spatially
separated components is formed. In both cases, these are closely
lying GaAs/AlGaAs double quantum wells, where either interwell
excitons are excited by a laser pulse [3, 4], or
two-dimensional electron layers are formed due to doping. In the
last case, the electron layers must be placed in a strong magnetic
field, normal to the layers, such that the total filling factor
should be $\nu_T = \nu_1 + \nu_2 = 1$ [5].  Since all these
systems have one and the same exciton mechanism for the interlayer
phase coherence [6], the physical properties of these systems in the
coherent state must be qualitatively similar.

    The present paper has mainly been stimulated by recent impressing
experiments of Spielman {\it et al.} [7, 8] who have found that if
a bilayer electron system transits into a phase-coherent state (in
which the quantum Hall effect is observed at $\nu_T = 1$), then
this transition is accompanied by a sharp rise in the differential
tunneling conductance $G_T$ at low interlayer voltages $V$. As the
temperature is lowered, this peak of tunneling conductance remains
of a finite height and width in contrast to the tunneling
conductance peak of the Josephson junction. In the parallel to the
layers magnetic field $H$ the peak height decreases the more
drastically, the higher is the field, and at $H
> 0.6$ T  the peak becomes practically indistinguishable.

    A number of papers have been devoted to theoretical interpretation of the
experimental results obtained by Spielman {\it et al.} For
example, Fogler and Wilczek [9] have treated the tunneling
conductance peak as a consequence of the Josephson effect in a
long inhomogeneous junction. In refs. [10, 11], the
interpretation of the peak has been based on the notion of the
finite time of phase coherence. Joglekar and MacDonald [12]
have performed both phenomenological and microscopic calculations
of tunneling conductance $G_T$ value at $V=0$. In ref. [13],
$G_T(V,H)$ was calculated using the phenomenological equation
similar to the Landau-Lifshitz equation for the magnetic moment.
Such a diversity of theoretical approaches in the interpretation
of experiment [7, 8] gives impetus to a consistent
microscopic consideration of the dynamics of phase-coherent
bilayer systems, this being just the subject of the present paper.
Though we consider the $n-p$ system in the absence of a
perpendicular magnetic field, the exciton nature of the collective
state in all the above-mentioned systems encourages us to believe
that the present results provide a qualitative description of
experiments by Spielman {\it et al.} [7, 8].

    An important but still not completely resolved problem for the systems with
electron-hole pairing is the problem of phase locking by interband
transitions [14], which  coincide with interlayer tunneling
transitions in the systems under consideration . The tunneling
transitions lift the degeneracy in the phase of the order
parameter, thereby locking the phase and making it equal to the
phase of tunneling matrix elements. The last statement is valid in
the absence of a parallel to the layers magnetic field. Kulik and
one of the present authors [15] have shown that in the
magnetic field parallel to the layers the phase is locked only in
the fields $H < H_{c1}$. (The critical field $H_{c1} \propto
|T_{12}|^{1/2}$, where $T_{12} (=|T_{12}|e^{i\chi})$ is the matrix element of
interlayer tunneling). At $H > H_{c1}$ the phase locking  is
lifted and the phase monotonically changes in the direction normal
to the field, this giving rise to spatial oscillations of
tunneling current (vortex state). The phase locking phenomenon
appears to exert an essential effect not only on the thermodynamic
properties of $n-p$ systems, but also on their kinetics.

    Relying on the microscopic approach, the present paper deals with the
response of phase-coherent $n-p$ system to the interlayer voltage
$V$. We demonstrate that similarly to the existence of the
critical field $H_{c1}$, in the case under consideration there
exists the threshold voltage $V_c(\propto |T_{12}|)$ that
quantitatively characterizes the degree of phase locking in the
$n-p$ system. At low voltages, $V < V_c$, the order parameter
phase is locked (steady-state), and the direct tunneling current
is proportional to $V$. The Ohmic character of a spatially uniform
tunneling current at $V < V_c$ means that in the phase-coherent
$n-p$ system there is no dc Josephson effect [16]. (The
absence of the dc Josephson effect in the two- layer {\it
electron} system has been established by Joglekar and MacDonald
[12]). At voltages $V > V_c$, the phase monotonically changes
with time, and this results in the tunneling current oscillations
with frequency $\omega = e\sqrt{V^2-V_c^2}$ (here $e$ is the
elementary charge, $\hbar = 1$) So, at $V > V_c$ the $n-p$ system
retains the essential feature of the ac Josephson effect in
superconductors, namely, the tunneling current oscillations at a
constant applied voltage. At the same time, the dissipative
character of the oscillating tunneling current (see below), the
nonuniversality of the voltage dependence of $\omega$, and the
presence of threshold voltage $V_c$ are specific to phase-coherent
bilayer $n-p$ systems.

    Further on, we show that the above-described "liberation" of the order
parameter phase at $V = V_c$ results in a sharp peak of $G_T(V)$,
the height of which is independent of $|T_{12}|$, and its width is
equal to $2 V_c$, i.e., for small $|T_{12}|$ the peak will be high
and sharp. Thus, in our opinion, the nature of the tunneling
conductance peak observed in the experiments of Spielman {\it et
al.} is closely connected with the phenomenon of order parameter
phase locking by tunneling transitions. The experimentally
observed suppression of $G_T(V)$ peak with an increasing parallel
magnetic field [8] also lends support in favor of this
interpretation, because, as indicated above, a sufficiently strong
in-plane magnetic field eliminates the phase locking.

    We are now coming to the analysis of the dynamics of a
phase-coherent $n-p$ system in the limit of a high pair density,
when the average distance between the electron-hole pairs is small
as compared to the characteristic pair size. The advantage of the
high-density limit lies in the possibility of considering the
phase-coherent system dynamics in the gapless state, when the gap
in the excitation spectrum becomes zero under the action of strong
depairing, and the order parameter $\Delta$ is reduced but remains
non-zero [17]. For the $n-p$ bilayer the order parameter is
proportional to the average $\langle \psi_1({\bf
r},t)\psi_2^+({\bf r},t)\rangle$, where $\psi_i^+({\bf r},t)$ is
the electron creation operator in the layer $i$. An essential
simplification consists in the fact that the absence of the gap
makes it possible to describe the dynamics of the phase-coherent
system only in terms of the complex order parameter
$(\Delta=|\Delta|e^{i\theta})$ without involving the dynamics of
the quasiparticle distribution function.

    The dynamic equation  for the
order parameter of the $n-p$ system was derived by the Green
function technique in our paper [18] and has the following
form:

\begin{equation}\label{1}
  -(\dot
\Delta-ieV\Delta)+\{A-B|\Delta|^2+D[{\frac{\partial}{\partial {\bf
r}}}+ {\frac{ie}{c}} ({\bf A}_1-{\bf A}_2)]^2\}\Delta +
{\frac{T_{12}}{\zeta \tau}}=0.
\end{equation}
The equation obtained is in a perfect agreement with the general
theory of a relaxation of an order parameter near a point of a
phase transition of the second kind (see, for instance [19]).
In accordance with this theory a state of a physical system under
a phase transition of the second kind can be described by an order
parameter, that is nonzero below  the transition point and  equal
to zero above this point. An equilibrium value of the order
parameter can be found from the condition that a variation of a
corresponding thermodynamic potential is equal to zero. In the
absence of  the interband hybridization the thermodynamic
potential for the condensate of electron-hole pairs with spatially
separated components can be presented in the form
\begin{equation}\label{p1}
  F=\int \Bigg\{ D |[-i\frac{\partial}{\partial{\bf
  r}}-\frac{e}{c}({\bf A}_1-{\bf A}_2)] \Delta|^2+
  A|\Delta|^2+\frac{1}{2}B |\Delta|^4\Bigg\} d {\bf r}
\end{equation}
Expression (\ref{p1}) is similar to the thermodynamic potential
for Cooper pairs in the Ginzburg-Landau theory, but here the term
$2e{\bf A}$ is replaced by the term $e({\bf A}_1-{\bf A}_2)$. Such
a modification is quite natural. Indeed, for the case of
electron-hole pairs with spatially separated components an
electron in the layer 1 "sees" the vector potential ${\bf A}_1$,
while a hole in the layer 2 "sees" the vector potential ${\bf
A}_2$. Since the signs of  the electron and hole charges are
different the vector potentials ${\bf A}_1$ and ${\bf A}_2$ are
subtracted from each other in eq. (\ref{p1}). In the equilibrium
the order parameter $\Delta({\bf r})$ is found from the condition
$\delta F/\delta\Delta({\bf r})=0$. At small deviation from the
equilibrium, when the derivative $\delta F/\delta\Delta({\bf r})$
is nonzero but a small one, the order parameter relaxation rate
(the derivative $\partial \Delta/\partial t$) is also small. In
the mean field approximation these two derivatives should be
proportional to each other. But it is necessary to take into
account that due to the gauge invariance of the theory the
derivative $\partial/\partial t$ can enter into the equation in a
combination with the term $ie(V_1-V_2)$, where $V_1$ and $V_2$ are
the electrostatic potentials in the layer 1 and 2,
correspondingly. As a result in the absence of the interband
hybridization one arrives to the equation (\ref{1}), where
$T_{12}=0$. In the presence of the interband hybridization the
Hamiltonian  of the system contains the terms linear in the order
parameter $\Delta$ and in the matrix elements $T_{12}$ and the
corresponding conjugate terms (and it means that the thermodynamic
potential contains the same terms). These terms play the role of a
source of electron-hole pairs. They are analogous to the terms
that appear in the Hamiltonian of a ferromagnet in an external
magnetic field. For the case of a magnet it results in an
appearance of a term linear in the magnetic field in the equation
for the order parameter. Since for the system considered the
matrix element $T_{12}$ is analogous to the magnetic field, a term
linear in $T_{12}$ should appear in  the equation for the order
parameter in the presence of the interband hybridization. We see
that eq. (\ref{1}) contains this term, indeed. The microscopic
analysis shows that in spite of the phenomenological arguments
presented look quite general, in reality,  eq. (\ref{1}) is valid
only in a rather narrow interval of the impurity concentration in
similarity with the Gor'kov-Eliashberg equation for the
superconductors with paramagnetic impurities [20].

 In the gapless situation under consideration, the
coefficients of the dynamic Ginzburg-Landau equation (\ref{1}) have the forms $
A(T)=(2\pi^2/ 3)\tau(T^2_c-T^2)$, $B=4m\tau/ 3 M$, $D=p_0^2\tau/
M^2$ [18]. Here $\tau$  is the electron elastic scattering
time ( for simplicity, it is considered equal to the hole elastic
scattering time), $T$ is the temperature ($k_B = 1$), $T_c$ is the critical
temperature, $M = m_1 + m_2$ is the pair mass, $m = {m_1 m_2/M}$
is the reduced mass of pair, $p_0$ is the Fermi momentum of
electrons and holes, $\zeta$ is the dimensionless constant of the
Coulomb interaction [2].
 It should be noted that eq. (\ref{1}) is derivated by expanding of the
anomalous Green function as a power series in $(\Delta/T_c)$ [18].
Since a term linear in the matrix element $T_{12}$ appears in
the expression for the order parameter, it is necessary that $|T_{12}|\ll T_c$
for the validity of eq. (\ref{1}).

    At low fields and currents, the modulus of the order parameter
varies only slightly in space and time. Assuming $|\Delta|$ to be
constant equal to $\Delta_0$, the imaginary part of eq. (\ref{1})
can be written as follows

\begin{equation}\label{2}
\dot\phi - D{\frac{\partial}{\partial {\bf
r}}}(\frac{\partial\phi}{\partial {\bf r}} - {\frac{2\pi d}{
\Phi_0}}[{\bf H} {\bf n}]) - eV + eV_c \sin\phi = 0 .
\end{equation}
 Here, the
gradient-invariant phase $\phi =\theta -\chi -(2\pi d /\Phi_0)A_z$
is introduced, $d$ is the interlayer distance, $\Phi_0 = hc/e$ is
the magnetic flux. The unit vector ${\bf n} =(0,0,1)$ is normal to
the layers and is directed from layer 1 (electron layer) to layer
2 (hole layer). The threshold voltage $V_c =
|T_{12}|/(e\zeta\tau\Delta_0)$.

    It is readily seen that in the uniform case eq. (\ref{2}) for the phase
$\phi$ is different from the equation $\dot\phi = eV$ that appears
in a number of papers and is treated as the Josephson relation for
phase-coherent bilayer systems. The occurrence of the term
proportional to $|T_{12}|$ in the dynamic equation for the phase
radically changes the solutions of this equation. Thus, in the
absence of external fields the stable steady-state and uniform
solution of eq. (\ref{2}) is $\phi = 0$, i.e., $\theta = \chi$,
and this means that the interlayer tunneling transitions hold the
order parameter phase locked. Below, we consider in detail how the
phase locking phenomenon influences the dynamic properties of
$n-p$ systems.

    We start from the analysis of dynamics of the $n-p$ system in the phase-coherent
state  for the spatially uniform case in the absence of the
magnetic field. Let the $n-p$ tunneling junction is incorporated
into the electrical circuit having the resistance $R$ and the
voltage source $\cal E$. The resulting voltage $V$ across the
$n-p$ tunneling junction determines the difference of
electrochemical potentials of the layers and thereby dictates the
carrier density in $n-$ and $p-$layers. If $\delta n$ is the
deviation of the electron density from the equilibrium one, then
the equality $eV = - \delta n/N_*(0)$ is valid, where the
renormalized density of states on the Fermi surface is $N_*(0) =
N(0)(1 +{\frac{e^2 m}{\pi C}})^{-1}$ ($C$ is the capacity of the
bilayer system related to the unit area; $N(0) = m/\pi$).

    In the approximation linear in $T_{12}$  the density of tunneling
current from layer 1 to layer 2 is equal to $J = J_c\sin\phi$,
where $J_c = 4eN(0)|T_{12}|\Delta_0/\zeta$ [18]. The charge
balance equation can be written as

\begin{equation}\label{3}
  eS\delta \dot n +{\frac{{\cal E} -V}{R}} - I_c\sin\phi = 0,
\end{equation}
where $S$ is the area of the $n-p$ junction, and $I_c =S J_c$.
Though below we assume ${\cal E} ={\rm const}$, it should be noted
that eq. (\ref{3}) also holds for the time-dependent voltage
source.

    Making use of the relationship between $ \delta n$ and $V$,
from eqs. (\ref{2}) and (\ref{3}) one can derive the second-order
equation for the phase $\phi$. In terms of dimensionless
variables, this equation takes on the following form, well known
in the theory of Josephson junctions
\begin{equation}\label{4}
\ddot\phi + \frac{1}{\sqrt\beta }(1+\epsilon\cos\phi )\dot\phi
+\sin\phi = \rho.
\end{equation}
 Here, the following dimensionless
parameters are introduced: $\beta = e {\cal E}_c t_0$, $\epsilon =
e V_c t_0$, $\rho = {\cal E}/{\cal E}_c$, where ${\cal E}_c = V_c
+ I_c R$, and $t_0 = e^2 N_*(0) R S$. The time is measured in the
$1/\omega_0$ units, where $\omega_0 = (e{\cal E}_c/t_0)^{1/2}$.

    Despite the coincidence of eq.(\ref{4}) with the dynamic equation for the phase
difference across the Josephson junction, the different meaning of
the parameter $\rho$ entering into these equations leads (as it
will be seen from what follows) to a substantially different
behavior of $n-p$ systems and Josephson junctions.

    A detailed analysis of dynamic states of the system described by
eq. (\ref{4}) was performed by Belykh {\it et al.} [21] Not
going into details of that analysis, we shall mention its main
results. For each value of the parameter $\epsilon $ one can find
the corresponding number $\beta_1$. At $\beta > \beta_1$ (large
resistances $R$), the range of $\rho$ values is split into three
adjacent intervals: $0 <\rho <\rho_c$, $\rho_c <\rho <1$ , $\rho
>1$ ($\rho_c(\beta,\epsilon)$ is the bifurcation value of the parameter $\rho$
[21]).
In the first interval, there
is only one stable solution $\phi=\arcsin\rho$; in the third
interval the only stable state is the limiting cycle embracing the
phase cylinder. In the intermediate (second) interval the both
solutions, $\phi=\arcsin\rho$  and the limiting cycle, are stable.
This non-uniqueness of the solution of eq. (\ref{4}) results in
the hysteresis of current-voltage characteristic (CVC) at $\beta >
\beta_1$. For $\beta<\beta_1$ (low resistances $R$) the stable
solutions will be $\phi=\arcsin\rho$ at $0 <\rho <1$ and the
limiting cycle at $\rho > 1$, while the interval $\rho$ with two
stable states drops out. Correspondingly, at $\beta<\beta_1$ the
CVCs have no hysteresis.

    Further on, we find the CVC and the differential tunneling conductance of
the $n-p$ system in a simple, but physically rather demonstrative,
case $R = 0$. In this limit, no distinction may be made between
$V$ and $\cal E$, $V_c$
 and ${\cal E}_c$, and the
dynamics of the system may be analyzed relying on eq. (\ref{2})
(without spatial derivatives). Since in the case considered we
have $\beta<\beta_1$, then the hysteresis of the CVC is absent.

If the system is spatially uniform and the voltage $V$
does not depend on the time the
equation (\ref{2}) can be integrated. One can see that for $V<V_c$
the equation  (\ref{2}) has the time independent solution
$\phi_0=\arcsin V/V_c$. In such a case the tunnel current, that does not depend
on time as well, is equal to $I_c \sin \phi_0=I_c V/V_c\equiv
V/R_c$. This current is proportional to the voltage $V$ applied
and it is a usual dissipative current.

 The corresponding tunneling conductance is given by

\begin{equation}\label{5}
  G_T =\frac{d I}{d V} =R_c^{-1} =4e^2 N(0) \tau \Delta_0^2 S.
\end{equation}
Note that at $V < V_c$  the tunneling conductance is constant and
is independent of the tunneling matrix element $|T_{12}|$ value.
This independence of tunneling conductance from $|T_{12}|$, and
also its proportionality to $\Delta_0^2(T)$ are in agreement with
the result of Joglekar and MacDonald [12] for $G_T$ at $V =
0$.

In case of $V>V_c$
 the integration yields the tunnel current equals to

\begin{equation}\label{d1}
  I(t)=2 I_c \frac{\tan \frac{\phi(t)}{2}}{1+\tan^2 \frac{\phi(t)}{2}}
\end{equation}
where
\begin{equation}\label{d2}
  \tan
  \frac{\phi(t)}{2}=V_c/V+\sqrt{1-(V_c/V)^2}\tan
  \left[\frac{e}{2}(V^2-V_c^2)^{\frac{1}{2}} (t-t_0)\right]
\end{equation}
One can see that the interlayer current oscillates with
the frequency $\omega=e (V^2-V_c^2)^{1/2}$ and this
current is not a sinusoidal one. Due to non-sinusoidal character
of the oscillations the average value of the tunnel current is
nonzero. The average current is the function on the voltage $V$.

\begin{equation}\label{6}
I = (I_c/V_c)(V-\sqrt {V^2 - V_c^2}).
\end{equation}

The behavior of the system considered is similar to the behavior
of a Josephson junction between two superconductors in a circuit,
where the junction is connected in series with a resistor and a
voltage generator. But in the case considered the essential
difference is that the resistor (with $R_c=V_c/I_c$) is embedded
in the junction and it cannot be deleted from the circuit. There
is not any transverse superconductivity in the systems considered.

Since according to eq. (\ref{6}) the tunneling current decreases
with an increasing voltage, the differential tunneling conductance
at $V > V_c$ is negative:

\begin{equation}\label{7}
  G_T(V) = -(I_c/V_c)[V(V^2 -V_c^2)^{-1/2} - 1].
\end{equation}

The conductance $G_T(V)$ has its maximum (constant) value at $|V|
< V_c$ and the discontinuity points at $V= \pm V_c$. At $|V| >
V_c$, as $|V|$ increases, the tunneling conductance monotonically
tends to zero, remaining negative. If we take into account the
fluctuation smoothing of the CVC, then the dependence of $G_T$ on
$V$ will look like a smooth curve with the maximum at $V = 0$
(approximately $2 V_c$ in width) and two minima at $V\approx\pm
V_c$. It is just this behavior of the $G_T(V)$ curve that was
observed in experiment [8] in the absence of the magnetic
field parallel to the layers.

    It should be noted that both at $V > V_c$ and $V < V_c$ the
spatially uniform tunneling current is dissipative. The reason for
the dissipation lies in the fact that the uniform interlayer
current causes the order parameter phase to deviate from its
equilibrium value and a continuous input of energy is required to
maintain this nonequilibrium state.

    Let now the bilayer $n-p$ structure be placed in a magnetic field
${\bf H}$ parallel to the layers and directed along the $x$-axis.
If $H > H_{c1} = (2\Phi_0/ \pi^2 d)(J_c M/ en_s)^{1/2}$ (the
two-dimensional density of the pairs $n_s = 4p_0^2 N(0)
(\tau\Delta_0)^2/ M$), then the magnetic field between the layers
has a nonuniform (vortex) component. We shall show that the CVCs
of the $n-p$ system in the magnetic field strongly differ from the
CVCs in the zero field and are substantially different at both low
and high resistances $R$. In the limiting case $R = 0$ (and $H \gg
H_{c1}$), the solution of eq. (\ref{2}) can be derived using the
perturbation theory. Putting $\phi = \phi_0 + \phi_1$, where
$\phi_0 = k y  + \omega t$ ( $k = 2\pi d H/\Phi_0$, $\omega = eV$)
and taking into account the correction term $\phi_1$ (proportional
to a small $T_{12}$ value) as a perturbation, we obtain the
following expression for the average tunneling current density:

\begin{equation}\label{8}
  J = J_c \frac{eV_c}{2}\frac{\omega}{(Dk^2)^2 + \omega ^2}.
\end{equation}
So, for $R =0$ the CVC has a wide diffusion maximum at $\omega =
Dk^2$.

    At high $R$ values, the charge transport from one layer to
the other over the electrical circuit is insignificant, and the
electron density dynamics in layer 1 is determined by the
continuity equation

\begin{equation}\label{9}
e\delta\dot n = {\rm div}_2{\bf j} + J_c\sin\phi,
\end{equation}
where ${\rm div}_2{\bf j}$ denotes the two-dimensional divergence
of the intralayer current ${\bf j} =- \frac{en_s}{
M}(\frac{\partial\phi}{\partial{\bf r}} - \frac{2\pi
d}{\Phi_0}[{\bf H}{\bf n}]).$ In the assumption that $eV \ll
\tau\Delta_0^2$, the above-described perturbation-theory procedure
yields the following equation for $\phi_1$:

\begin{equation}\label{10}
\ddot\phi_1 - D\frac{\partial^2\dot\phi_1}{\partial^2{\bf r}} -
u_0^2\frac{\partial^2\phi_1}{\partial^2{\bf r}} = - \frac{J_c}{
eN_*(0)} \sin\phi_0,
\end{equation}
where $u_0 = (n_s/MN_*(0))^{1/2}$. Unlike the $R = 0$ case, the
left-hand part of eq. (\ref{10}) has a wave character rather than
a diffusion character. Correspondingly, the expression for the
average tunneling current density

\begin{equation}\label{11}
J = J_c\frac{1}{2 \lambda_J^2}\frac{\omega\alpha k^2}{
(\omega^2/u_0^2-k^2)^2 + (\omega\alpha k^2)^2}
\end{equation}
has the resonance at $\omega = u_0 k$, the width of which is
determined by the attenuation $\alpha = D/u_0^2$. This resonance
results from the coincidence between the plasmon velocity $u_0$ in
the bilayer structure and the velocity of the magnetic-field
vortices. The parameter $\lambda_J$ equals $(e n_s/M J_c)^{1/2}$.

From relations (\ref{8}) and (\ref{11}) it follows that at $H \gg
H_{c1}$ the $G_T(0)$ value is proportional to a small $|T_{12}|^2$
value, i.e., the differential tunneling conductance peak
(occurring at $H=0$) is strongly suppressed. The reason for this
suppression lies in the fact that at $H > H_{c1}$ the phase $\phi$
monotonically varies with the coordinate, and in this case eq.
(\ref{2}) has no stationary solution at finite voltage, i.e., no
phase locking arises.

    Thus, the present work has demonstrated in the frame of the consistent
microscopic approach that in phase-coherent bilayer $n-p$ systems
the known phenomenon of order parameter phase locking by matrix
elements of tunneling $T_{12}$ leads to a sharp peak in
differential tunneling conductance $G_T(V)$ at $V = 0$. The peak
height is independent of $|T_{12}|$ and its width is proportional
$|T_{12}|$, i.e., at weak tunneling the peak is high and sharp.
These results are in qualitative agreement with the peculiarities
of $G_T(V)$ observed in electron bilayer systems in the regime of
an integral quantum Hall effect at the total filling factor $\nu_T
= 1$. We stress once again that though the theory developed here
describes the $n-p$ system without a transverse magnetic field,
the present results are in qualitative agreement with the data
from experiments on electron bilayer systems in a strong
transverse magnetic field. This agreement does not seem to be
accidental. The reason is that most likely the strong magnetic
field does not affect the structure of the equation that defines
the  dynamics of the order parameter, but only changes the values
of coefficients entering into this equation.

    This work was supported by the INTAS program, grant No 01-2344.

\pagebreak
\centerline{ References}
\vskip 1 cm

1. Yu.E. Lozovik and V.I. Yudson, Zh. Eksp. Teor. Fiz {\bf 71}, 738 (1976) [
Sov. Phys. JETP {\bf 44}, 389 (1976)].

2. S.I. Shevchenko, Fiz. Nizk. Temp.{\bf 2}, 505 (1976)
[Sov. J Low Temp. Phys. {\bf 2}, 251 (1976)].

3. L.V. Butov, C.W. Lai, A.L. Ivanov, A.C. Gossard, and D.S. Chemia,
Nature, {\bf 417}, 47 (2002); L.V. Butov, A.C. Gossard, and D.S. Chemia,
Nature, {\bf 418}, 751 (2002).

4. A.V. Larionov, V.B. Timofeyev, J. Hvam, K. Soerensen, Pis'ma v
ZhETF {\bf 75}, 233 (2002) [JETP Letters {\bf 75}, 200 (2002)];
A.A. Dremin, V.B. Timofeyev, A.V. Larionov, J. Hvam, K. Soerensen,
Pis'ma v ZhETF, {\bf 76}, 526 (2002) [JETP Letters {\bf76}, 450 (2002)].

5. E.g., see J.P. Eisenstein in {\it Perspectives in Quantum Hall
Effects}  edited by S. Das Sarma and A. Pinczuk (Wiley, New York, 1997).

6. A.H. MacDonald and E.H. Rezayi, Phys. Rev. B {\bf 42}, 3224 (1990).

7. I.B. Spielman, J.P. Eisenstein, L.N. Pfeiffer, and K.W. West, Phys. Rev. Lett.
{\bf 84}, 5808 (2000).

8. I.B. Spielman, J.P. Eisenstein, L.N. Pfeiffer, and K.W. West, Phys. Rev. Lett.
{\bf 87}, 036803 (2001).

9. M.M. Fogler and F. Wilczek, Phys. Rev. Lett. {\bf 86}, 1833 (2001).

10. L. Balents and L. Radzikhovsky, Phys. Rev. Lett. {\bf 86}, 1825 (2001).

11. A. Stern, S.M. Girvin, A.H. MacDonald, and N. Ma, Phys. Rev. Lett. {\bf 86},
1829 (2001).

12. G.N. Joglekar and A.H. MacDonald, Phys. Rev. Lett. {\bf 87}, 196802 (2001).

13. M. Abolfath, R. Khomeriki, and K. Mullen, cond-mat/0208236.

14. R.R. Gusejnov and L.V. Keldysh have first demonstrated that the interband
transitions lift the phase degeneracy of the wave function of the electron-
hole condensate in the exciton dielectric: R.R. Gusejnov and L.V. Keldysh,
Zh. Eksp. Teor. Fiz. {\bf 63}, 2255 (1972) [Sov. Phys. JETP {\bf 36},
1193 (1972)].

15. I.O. Kulik and S.I. Shevchenko, Fiz. Nizk. Temp. {\bf 2}, 1406 (1976)
[Sov. J. Low Temp. Phys. {\bf2},687 (1976)].

16. S.I. Shevchenko, "Theory of low dimensional superfluidity in Bose- and
Fermi-systems", thesis, Kharkov, 1992 (unpublished).

17. Similarly to magnetic impurities in superconductors, the usual
(nonmagnetic) impurities and
crystal lattice distortions in the n-p system  suppress the
order parameter and lead to the transition of the system to the gapless
state in the
narrow range of defect concentrations in the vicinity of the critical concentration.
E.g., see A.I. Bezuglyj and S.I. Shevchenko, Fiz. Nizk. Temp.
{\bf 3}, 428 (1977) [Sov. J. Low Temp. Phys. {\bf 3}, 204 (1977)].

18. A.I. Bezuglyj and S.I. Shevchenko, Fiz. Nizk. Temp. {\bf 4}, 454 (1978)
[Sov. J Low Temp. Phys. {\bf 4}, 222 (1978)].

19. L.D. Landau, E.M. Lifshitz, {\it Physical Kinetics}, Course in Theoretical
Physics Vol. 10 (Nauka, Moscow, 1979).

20. L.P. Gor'kov, G.M. Eliashberg, Zh. Eksp. Teor. Fiz. {\bf 54}, 612 (1968).

21. V.N. Belykh, N.F. Pedersen, and O.H. Soerensen, Phys. Rev. {\bf B16}, 4853
(1977).

\end{document}